\RequirePackage{lineno}
\documentclass[aps,showpacs,floatfix,nofootinbib,twocolumn]{revtex4}
\usepackage{CJKutf8}
\usepackage{color,amsmath,amssymb,graphicx,latexsym,subfigure}
\usepackage{threeparttable,multirow,txfonts,ulem,url,float}

\def\GEANT{G{\footnotesize EANT}4 }

\newcommand{\chname}[1] {\begin{CJK}{UTF8}{gbsn}{#1}\end{CJK}}

\begin{document}

\title{Comparison of proton shower developments in the BGO calorimeter of the Dark Matter Particle Explorer between GEANT4 and FLUKA simulations}

\author{Wei Jiang(\chname{蒋维})$^{1,2}$}
\author{Chuan Yue(\chname{岳川})$^1$\footnote{Corresponding author: yuechuan@pmo.ac.cn}}
\author{Ming-Yang Cui(\chname{崔明阳})$^1$\footnote{Corresponding author: mycui@pmo.ac.cn}}
\author{Xiang Li(\chname{李翔})$^{1}$}
\author{Qiang Yuan(\chname{袁强})$^{1,2}$}
\author{Francesca Alemanno$^{3,4}$}
\author{Paolo Bernardini$^{5,6}$}
\author{Giovanni Catanzani$^{7,8}$}
\author{Zhan-Fang Chen(\chname{陈占方})$^{1,2}$}
\author{Ivan De Mitri$^{3,4}$}
\author{Tie-Kuang Dong(\chname{董铁矿})$^{1}$}
\author{Giacinto Donvito$^{9}$}
\author{David Francois Droz$^{10}$}
\author{Piergiorgio Fusco$^{9,11}$}
\author{Fabio Gargano$^{9}$}
\author{Dong-Ya Guo(\chname{郭东亚})$^{12}$}
\author{Dimitrios Kyratzis$^{3,4}$}
\author{Shi-Jun Lei(\chname{雷仕俊})$^{1}$}
\author{Yang Liu(\chname{刘杨})$^{1}$}
\author{Francesco Loparco$^{9,11}$}
\author{Peng-Xiong Ma(\chname{马鹏雄})$^{1,2}$}
\author{Giovanni Marsella$^{6,12}$}
\author{Mario Nicola Mazziotta$^{9}$}
\author{Xu Pan(\chname{潘旭})$^{1,2}$}
\author{Wen-Xi Peng(\chname{彭文溪})$^{13}$}
\author{Antonio Surdo$^{6}$}
\author{Andrii Tykhonov$^{10}$}
\author{Yi-Yeng Wei(\chname{魏逸丰})$^{14}$}
\author{Yu-Hong Yu(\chname{余玉洪})$^{15}$}
\author{Jing-Jing Zang(\chname{藏京京})$^{1}$}
\author{Ya-Peng Zhang(\chname{张亚鹏})$^{15}$}
\author{Yong-Jie Zhang(\chname{张永杰})$^{15}$}
\author{Yun-Long Zhang(\chname{张云龙})$^{14}$}

\affiliation{
$^1$Key Laboratory of Dark Matter and Space Astronomy, Purple Mountain Observatory, Chinese Academy of Sciences, Nanjing 210023, China\\
$^2$School of Astronomy and Space Science, University of Science and Technology of China, Hefei 230026, China\\
$^3$Gran Sasso Science Institute (GSSI), Viale F. Crispi 7, I-67100, L'Aquila, Italy\\
$^4$Istituto Nazionale di Fisica Nucleare, Laboratori Nazionali del Gran Sasso, Via G.Acitelli 22, I-67100, Assergi, L'Aquila, Italy\\
$^5$Dipartimento di Matematica e Fisica E. De Giorgi, Universit\`a~del Salento, I-73100 Lecce, Italy\\
$^6$Istituto Nazionale di Fisica Nucleare (INFN)--Sezione di Lecce, I-73100 Lecce, Italy \\
$^7$INFN Section of Perugia, I-06100 Perugia, Italy \\
$^8$University of Perugia, I-06100 Perugia, Italy \\
$^9$Istituto Nazionale di Fisica Nucleare (INFN)--Sezione di Bari, I-70125, Bari, Italy \\
$^{10}$Department of Nuclear and Particle Physics, University of Geneva, CH-1211, Switzerland\\
$^{11}$Dipartimento Interateneo ``M. Merlin'' dell'Universit\`a~degli Studi di Bari e del Politecnico di Bari,  I-70125, Bari, Italy \\
$^{12}$Dipartimento di Fisica e Chimica ``E. Segr\`e'', via delle Scienze Edificio 17, Universit\`a Degli Studi di Palermo, I-90128, Palermo, Italy\\
$^{13}$Institute of High Energy Physics, Chinese Academy of Sciences, YuquanLu 19B, Beijing 100049, China\\
$^{14}$State Key Laboratory of Particle Detection and Electronics, University of Science and Technology of China, Hefei 230026, China\\
$^{15}$Institute of Modern Physics, Chinese Academy of Sciences, Nanchang Road 59, Lanzhou 730000, China\\
}

\begin{abstract}
The DArk Matter Particle Explorer (DAMPE) is a satellite-borne detector 
for high-energy cosmic rays and $\gamma$-rays. To fully understand the
detector performance and obtain reliable physical results, extensive
simulations of the detector are necessary. The simulations are 
particularly important for the data analysis of cosmic ray nuclei,
which relies closely on the hadronic and nuclear interactions of particles 
in the detector material. Widely adopted simulation softwares include
the \GEANT and FLUKA, both of which have been implemented for the DAMPE simulation
tool. Here we describe the simulation tool of DAMPE and compare the results 
of proton shower properties in the calorimeter from the two simulation 
softwares. Such a comparison gives an estimate of the most significant
uncertainties of our proton spectral analysis.
\end{abstract}

\date{\today}

\maketitle

\section{Introduction}
The magnetic spectrometer experiments such as PAMELA and AMS-02 have
pushed the precise measurements of energy spectra of cosmic rays (CRs) 
to rigidities of $\sim$TV (e.g., \cite{Adriani:2011cu,Aguilar:2015ooa,
Aguilar:2017hno}). At higher energies, the measurements still have 
large uncertainties, which hinder a better understanding of the origin 
and propagation of CRs \cite{Tanabashi:2018oca}. In recent years, a
number of space calorimeter experiments have been launched, such as the
CALET \cite{Torii:2019wxn}, NUCLEON \cite{Atkin:2015nrd}, 
DAMPE \cite{ChangJ2014,DmpMission}, and ISS-CREAM \cite{Kang:2019pfl},
which have already or are expected to improve the direct measurements
of CR spectra remarkably.

The Dark Matter Particle Explorer,
is the first Chinese satellite for astroparticle physics studies.
It was launched on December 17, 2015, and has operated in a sun-synchronous 
orbit for more than 4 years ever since. The DAMPE is dedicated to 
indirectly detect the annihilation or decay products of dark matter 
via high-energy-resolution measurements of CR electrons plus positrons 
and $\gamma$-rays. As a CR particle detector, the DAMPE can also explore 
the origin of CRs, as well as the transient high-energy $\gamma$-ray
sky \cite{DmpMission,Yuan:2018rys}. 

The DAMPE detector is made up of four sub-detectors, including a Plastic 
Scintillator Detector (PSD; \cite{YuYH2017}), a Silicon Tungsten 
tracKer-converter (STK; \cite{Azzarello2016}), a Bismuth Germanium Oxide 
imaging calorimeter (BGO; \cite{ZhangZY2016}), and a NeUtron Detector
(NUD; \cite{HuangYY2020}). These four sub-detectors cooperate 
to give high-precision measurements of the charge, direction, energy, and 
identity of each incident particle (see Ref.~\cite{DmpMission} for more
details). The on-orbit calibration shows that the detector is quite
stable with time after the launch \cite{DmpCalibration}. Up to now,
high-precision measurements of the CR electron plus positron spectrum
and the proton spectrum in wide energy ranges have been reported by
the DAMPE collaboration \cite{DmpElectron,DmpProton}. 

Dedicated Monte Carlo (MC) simulations of the particle response in the 
DAMPE detector, including the impacts of the modules of the satellite 
platform, are important for understanding the detector performance, such 
as the evaluations of efficiencies, the energy and direction responses,
and the background contaminations. For the hadronic CR analysis
simulations are even more crucial since the calorimeter only records
a fraction of the particle's energy and the full energy response can
only be obtained by simulations. Two of the leading softwares widely
used for particle simulations are \GEANT \cite{GEANT4} and 
FLUKA \cite{FLUKA2005,FLUKA}. \GEANT is a C++ toolkit to simulate the passage 
of particles through matter. It has a large set of physics processes 
handling the complicated interactions of particles in the matter up 
to 100 TeV energies. FLUKA is a FORTRAN based, fully integrated 
particle physics simulation package for calculations of particle 
transport and interactions with matter in the energy range from MeV up to PeV.


The hadronic showers are essentially hybrid cascades of hadronic processes
and electromagnetic processes. The inelastic hadronic interactions produce
secondary particles (mainly pions), and charged pions may induce
additional hadronic interactions, while the neutral pions would most 
likely decay into photons which experience electromagnetic cascades 
further.  
Other physical processes governing the hadronic showers
include nuclear fragmentation, ionization, elastic scattering, nuclear 
de-excitation, and so on \cite{Tanabashi:2018oca}. 
The HARP-CDP experiments reported the comparison of the production yields of 
the interactions of protons and charged pions with beryllium, copper, and
tantalum nuclei between these two software tools with the momentum up to 15 GeV/c \cite{ComparisonGF}.
A poor agreement between the \GEANT QGSP\_BERT \cite{GEANT4} and FLUKA was presented.
Recent development of \GEANT \cite{GEANT4_2016} implemented the Fritiof(FTF) \cite{FTFP_G4_1,FTFP_G4_2}
model to simulate the inelastic hadron-nucleus processes over the energy range up to 100 TeV \cite{GEANT4_PhLG}.
FLUKA can simulate the interaction and propagation in matter of many species of particles with high accuracy, 
especially hadrons of energies up to 20 TeV (up to 10 PeV when it is interfaced with the DPMJET3 code \cite{dpmjetfluka}) 
and all the corresponding antiparticles, neutrons down to thermal energies and heavy ions \cite{FLUKA}.

The hadronic cascade processes have relatively large fluctuations, resulting in
relatively large energy dispersion. Usually the spectral unfolding
method is necessary for the reconstruction of energy spectra of
CR nuclei, which depends on the MC simulations. In this work, 
we carry out a comparing study of the proton shower behaviors obtained
from the \GEANT and FLUKA simulations of the DAMPE detector. Their
difference can be considered as an estimate of the systematic uncertainties of the
proton spectrum measurements \cite{DmpProton}.

\section{Detector simulations}

\subsection{Geometry configuration}
\begin{figure}[ht]
\centering
\includegraphics[width=0.50\textwidth]{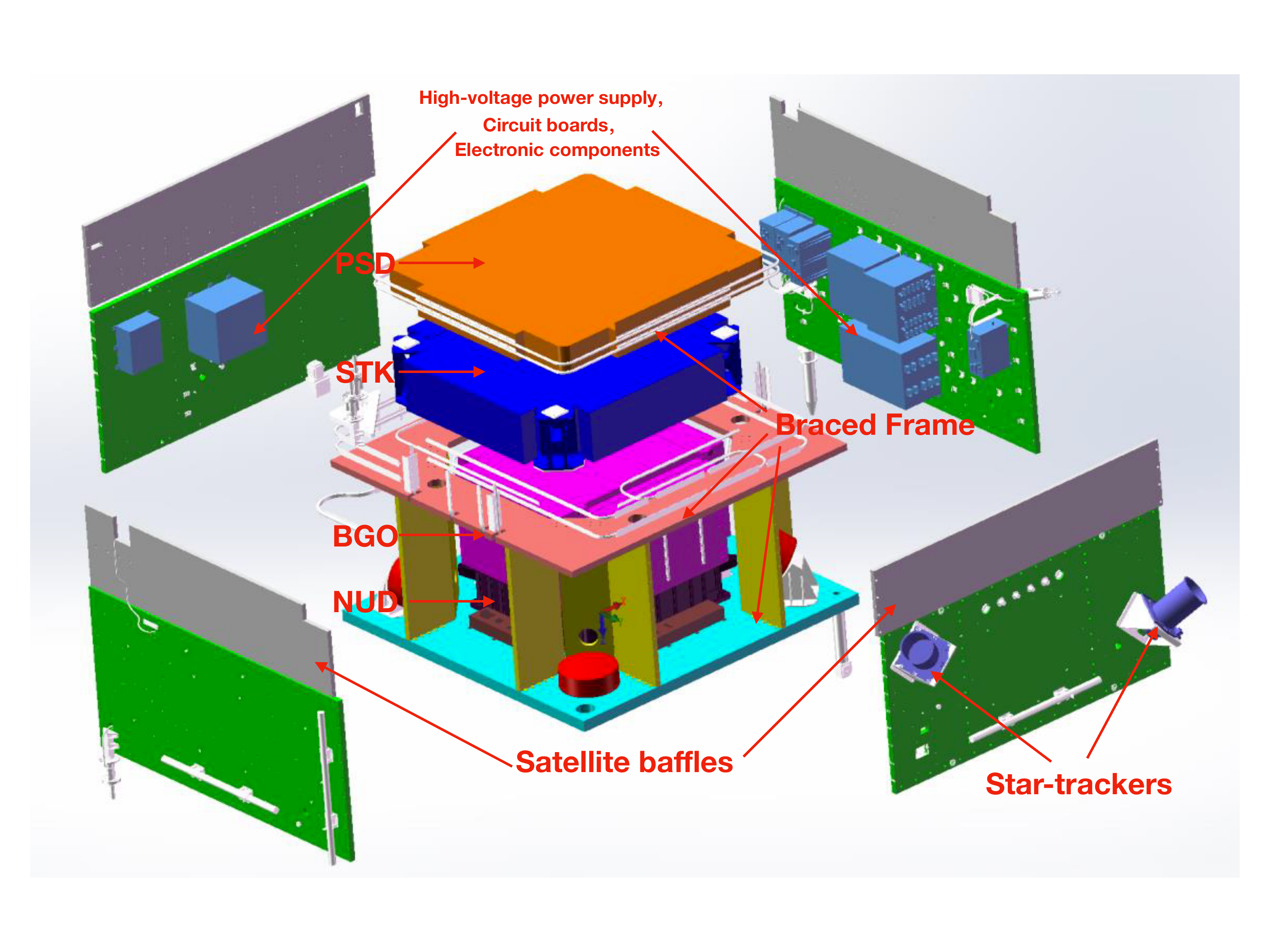}
\caption{The geometric model of the entire satellite, including the detector
payload and the satellite platform.}
\label{fig::geometry}
\end{figure}

Both \GEANT and FLUKA simulations are based on an accurate geometric model 
including both the payload and the satellite platform,
which is designed to study various characteristics and performance improvement of the detector.
The sizes, shapes, positions of all components of the satellite in the designed
geometry are measured and validated  in detail during the assembly of the satellite.
Meanwhile, the materials information of all the parts is appended into the geometry.
We take into account accurate atomic composition of different elements in 
the detector for precise hadronic and electromagnetic shower cascade simulation.
For most parts of the satellite, the manufacturers provide detailed element components.
For the remaining filling materials and electronic components whose compositions were unknown,
we send their samples to analytical laboratories for detailed measurements to get the exact mass fractions of atoms. 
Thus, the geometric model of the entire satellite is established, which accurately describes the detection units, the supporting structure and 
 the filling cushioning materials of the sub-detectors, as well as the frame and electronic components of the satellite, as shown in Fig.~\ref{fig::geometry}. 

\begin{figure}[htbp]
\centering
\includegraphics[width=0.5\textwidth]{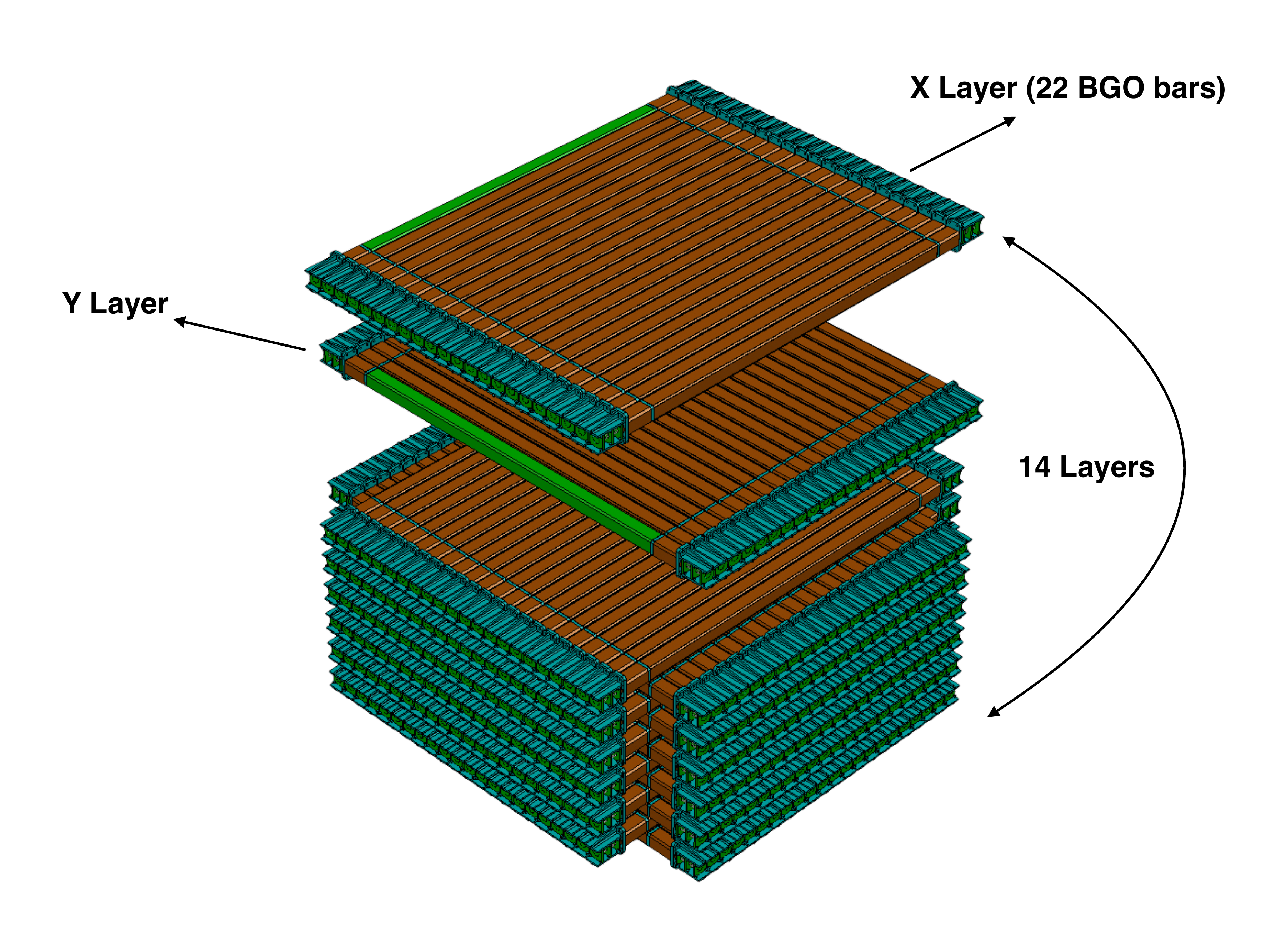}
\caption{The geometric model of the BGO calorimeter.}
\label{fig::bgo_geo}
\end{figure}

The BGO is the core sub-detector of the satellite payload. The characteristics of its interaction with protons are emphasized in this paper.
The detailed structure of the BGO is shown in Fig.~\ref{fig::bgo_geo}.
We precisely configure the geometry of the BGO which consists of 14 layers, each with 22 crystal bars.
Each bar is an independent detection unit with independent readout circuits at the two ends, assembled in a braced frame with cushioning material filling all the internal gaps. 
 After a great deal of measurements and verifications, we configure a precise BGO model in the geometry.

Following the configuration and validation of the designed geometry, we call 
the Geometry Description Markup Language (GDML) \cite{GDML} interface
for the specific program implementation of \GEANT simulation, called the ``GDML geometry'.
The parameters of the GDML geometry are directly derived from the designed documents
and measured results and are checked repeatedly to confirm that this geometry reflects 
the real situation of the satellite accurately. 
The  GDML geometry is integrated into the DAMPE offline software framework \cite{DMPSW},
which performs a series of standardized procedures (calibration, reconstruction and analysis)
from the real ``raw data'', the original signal collected by each prob cell of DAMPE, to scientific results. 
Therefore, the GDML geometry is applied as a unified interface for the data analysis of DAMPE.
On the other hand, the FLUKA simulation is fully integrated and closed source code. 
It only allows the Combinatorial Geometry (CG) \cite{MORSE} interface to develop the geometry.
The geometry is rewritten from the GDML for the FLUKA simulation, called the ``CG geometry''.
After careful and repeated checks of these two geometry models in every detail,
we are confident that they are identical between each other and consistent with the real satellite geometry,
although there are slight differences in some micro components of the satellite which
are negligible during the simulation.

\subsection{Data process}
\begin{figure}[htbp]
\centering
    \includegraphics[width=0.5\textwidth]{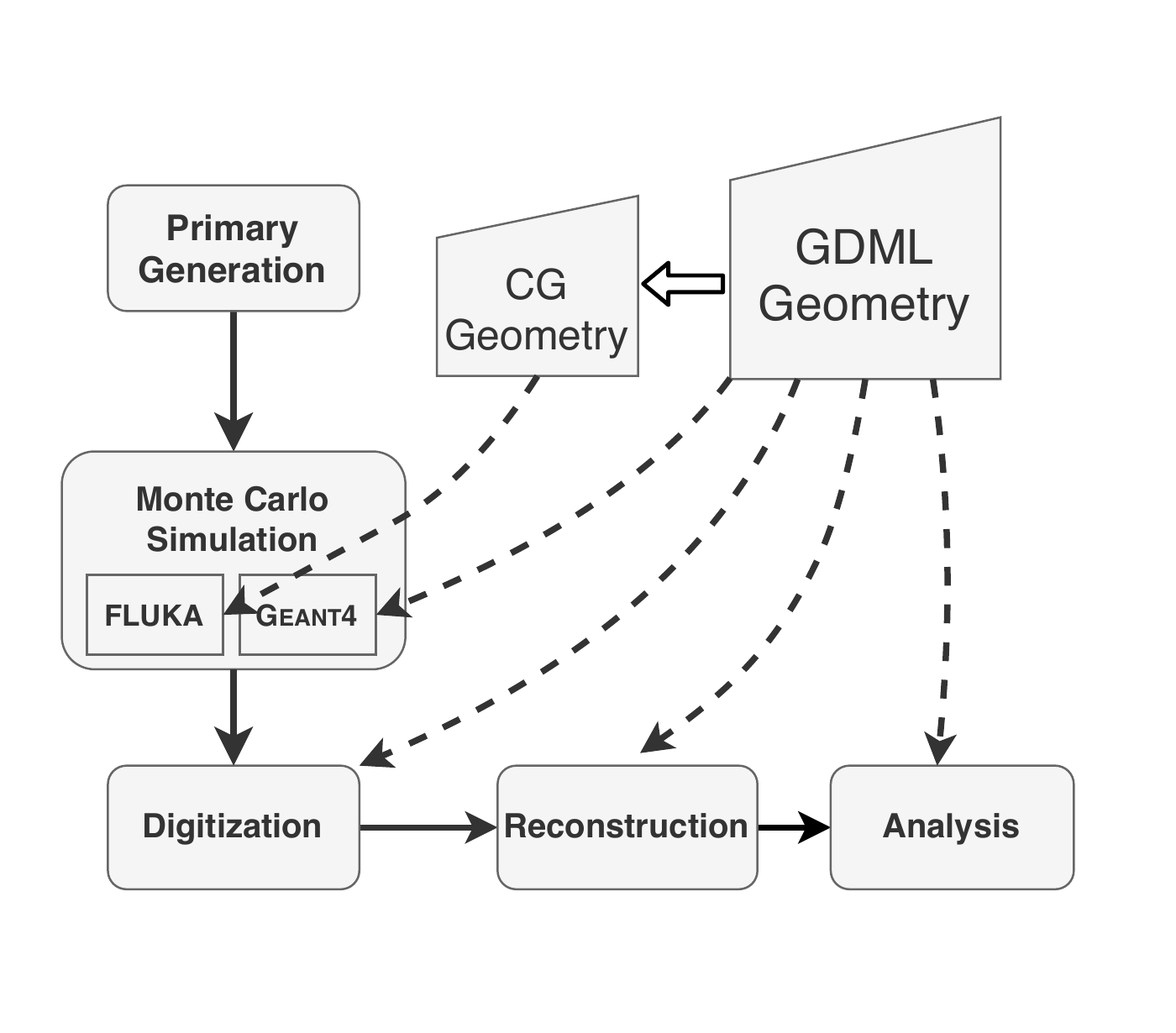}
\caption{General scheme for the full simulation data process.}
\label{fig::simu_setups}
\end{figure}

The data flow for the complete simulation process is shown in Fig.~\ref{fig::simu_setups}, including primary generation, MC simulations, digitization, reconstruction and analysis. 
The first package, primary generation, creates the incident particles feeding the MC simulation including  various distributions of incident positions, directions and energies.
In this work, we generate a flux of primary protons distributed isotropically 
with a single power-law energy spectrum with the index -1 from 10 GeV to 100 TeV.
The \GEANT simulation package is integrated into the DAMPE offline software framework
so that the \GEANT simulation can be managed and coordinated as standardized configuration.
We choose the FTFP\_BERT \cite{GEANT4_2016} from the \GEANT Physics Lists \cite{GEANT4_PhLG} 
with the default configuration recommended in the \GEANT documents \cite{GEANT4_PhRM}.
The FLUKA simulation runs in a separate operating environment, and is performed with the following settings:
\begin{itemize}
\item the {\tt PEANUT} package is activated in the whole energy range for any reaction;
\item the minimum kinetic energy for {\tt DPMJET-III} is set to $5 ~  \mathrm{GeV}/n $ (applying only to reactions between two nuclei heavier than a proton);
\item the minimum kinetic energy for {\tt RQMD} is set to $0.125 ~ \mathrm{GeV}/n$ (applying only to reactions between two nuclei heavier than a proton);
\item the same output format as the \GEANT simulation. 
\end{itemize}

Following numerous tests and validations to the developed simulation package including a set of algorithms which are responsible for generating the interactions of particles with the detector based on both the G{\footnotesize EANT}4.10.03\footnote{http://geant4.web.cern.ch} and FLUKA 2011.2x\footnote{https://www.fluka.org}, we allocate massive computing resources to run these programs, producing the simulation data of billions of protons.

Then, we run the digitization package to convert the physical information into the digital signal of each detection unit assigning a digital ID.
In such a way the digital information of the simulation is in the same format as the real ``raw data''.   
Accordingly, we can run the reconstruction package which contains 
large amounts of code for the for building up the physical signals 
including deposited energy, reconstructed tracks and charge of each event from the ``raw data''. 
This package is organized as a series of algorithms that act 
successively to process the on-orbit data on a daily basis \cite{DmpCalibration}. 
The massive code to obtain the scientific results and the instrument performance of the detector based on the reconstructed data is collectively referred to as the analysis package,
which is the result of collective efforts of many researchers.
All the code in the package undergoes continues enhancement and version update
as the detector comprehension improves with time.
Major published results were also obtained using the package to analyze the on-orbit data and simulation data. 
In these analysis packages, the event selection packages including a list of selection conditions for target particles 
are fundamental for the analyses.
In this work, we focus on figuring out some features to analyze the response of protons in the BGO calorimeter.
All the below results are obtained based on the selected proton samples
following the event selections in Ref.~\cite{DmpProton}.

\section{Results}

\subsection{Tigger efficiency}
Firstly, we investigate the tigger efficiencies for simulations using \GEANT and FLUKA.
DAMPE has four different triggers implemented on orbit: the Unbiased trigger,
the Minimum Ionizing Particle (MIP) trigger, the Low-Energy (LE) trigger,
and the High-Energy (HE) trigger \cite{ZhangYQ2019}. The Unbiased and MIP triggers
are designed for the detector calibration \cite{DmpCalibration}, while the LE and HE triggers 
correspond to low threshold and high threshold triggering signals respectively.
In the proton analysis, the events are required to meet the HE trigger condition in order to
guarantee that the shower development starts above or at the top of the calorimeter.
The HE trigger efficiency is one of the most important factors related to
the effective acceptance estimation. For different hadronic integration models, 
the shower start-point and the secondaries from the first inelastic interaction 
would be different. As a result, we would consider the difference of
the HE trigger efficiencies between \GEANT and FLUKA simulations.

\begin{figure}[htbp]
  \centering
  \includegraphics[width=.48\textwidth]{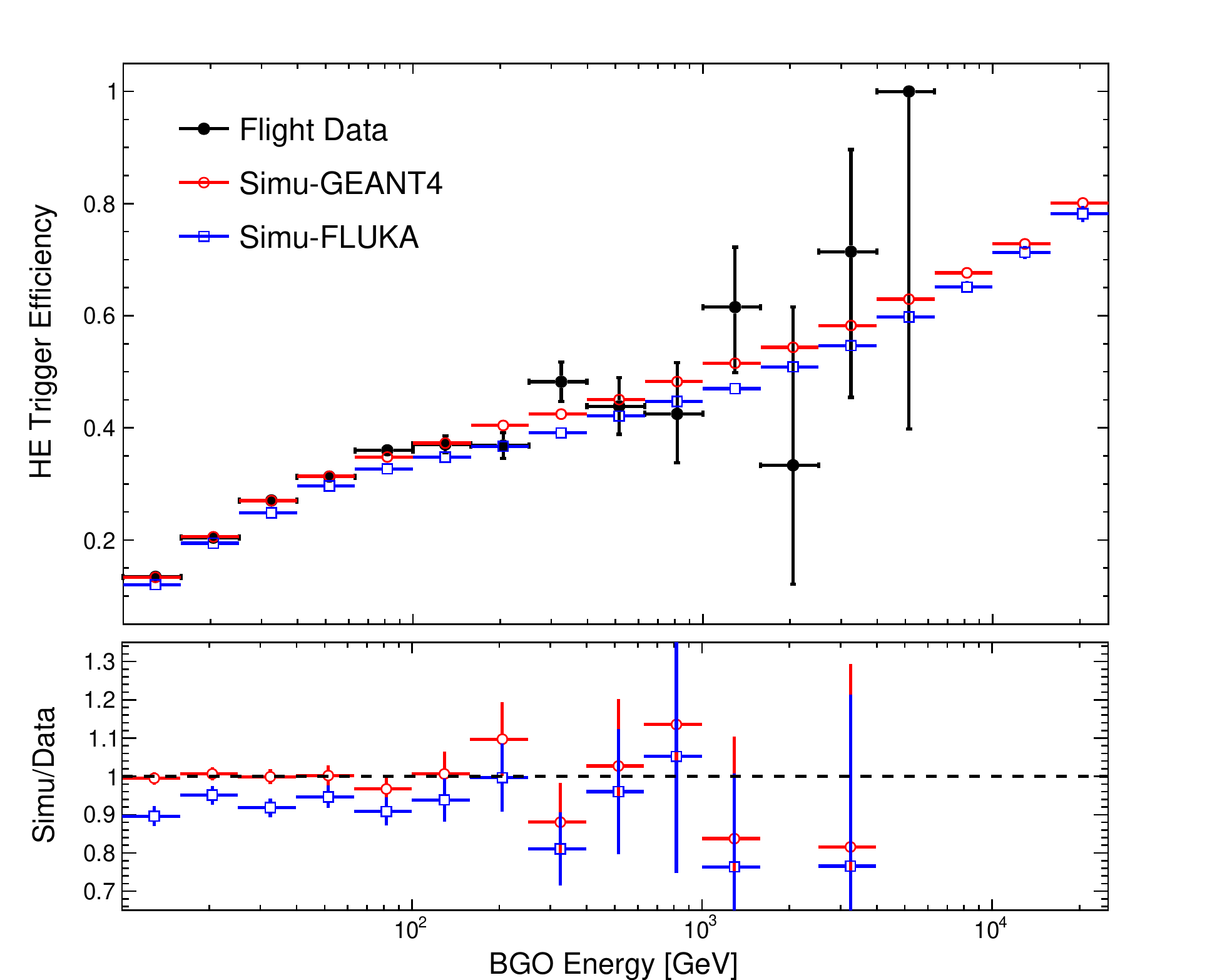}
  \caption{{\bf The HE tigger efficiencies for protons from FLUKA, \GEANT and flight data.} 
  The top panel shows the HE tigger efficiencies defined by Eq. \ref{eq::eff}. 
  The bottom panel shows the efficiency ratio of FLUKA and \GEANT to the flight data.
    }
  \label{fig::trigger}
\end{figure}

The HE trigger efficiency is estimated by means of the Unbiased trigger samples. 
The Unbiased trigger events are pre-scaled by a factor of 512 at latitudes $\le 20^{\circ}$ 
and 2048 at latitudes $> 20^{\circ}$. The HE trigger efficiency for protons is computed as 
\begin{equation}
\varepsilon_{\rm trigger} =\frac{N_{\rm HE \& Unb}}{ N_{\rm Unb}}, 
 \label{eq::eff}
\end{equation}
where $N_{\rm Unb}$ is the number of proton events passing the Unbiased 
trigger condition and $N_{\rm HE \& Unb}$ is the number of ones which 
both pass the HE and Unbiased trigger conditions. 
Fig. \ref{fig::trigger} shows the comparison of HE trigger efficiencies 
among the flight data, \GEANT and FLUKA. Despite of the limited statistics 
of flight data, it suggests that the \GEANT achieves a good agreement with 
flight data in the whole energy range, while FLUKA presents a systematic 
deviation of $\sim-5\% $ compared with the \GEANT and flight data .

\subsection{Total energy deposit}
The energy of an incident proton is measured by the sum of energy deposits of all BGO crystals in the calorimeter, i.e. the total energy deposit. 
Due to the limited vertical thickness of the BGO calorimeter ($\sim1.6$ nuclear interaction length) and the missing energy due to muon and neutrino components in hadronic showers, the total energy deposit measured by DAMPE would underestimate the intrinsic kinetic energy of an incident proton. 
In order to deconvolute the measured deposit spectrum into the initial spectrum, we need a good knowledge of the energy response, which is determined by the MC simulations. 
\begin{figure}[htbp]
  \centering
  \includegraphics[width=.48\textwidth]{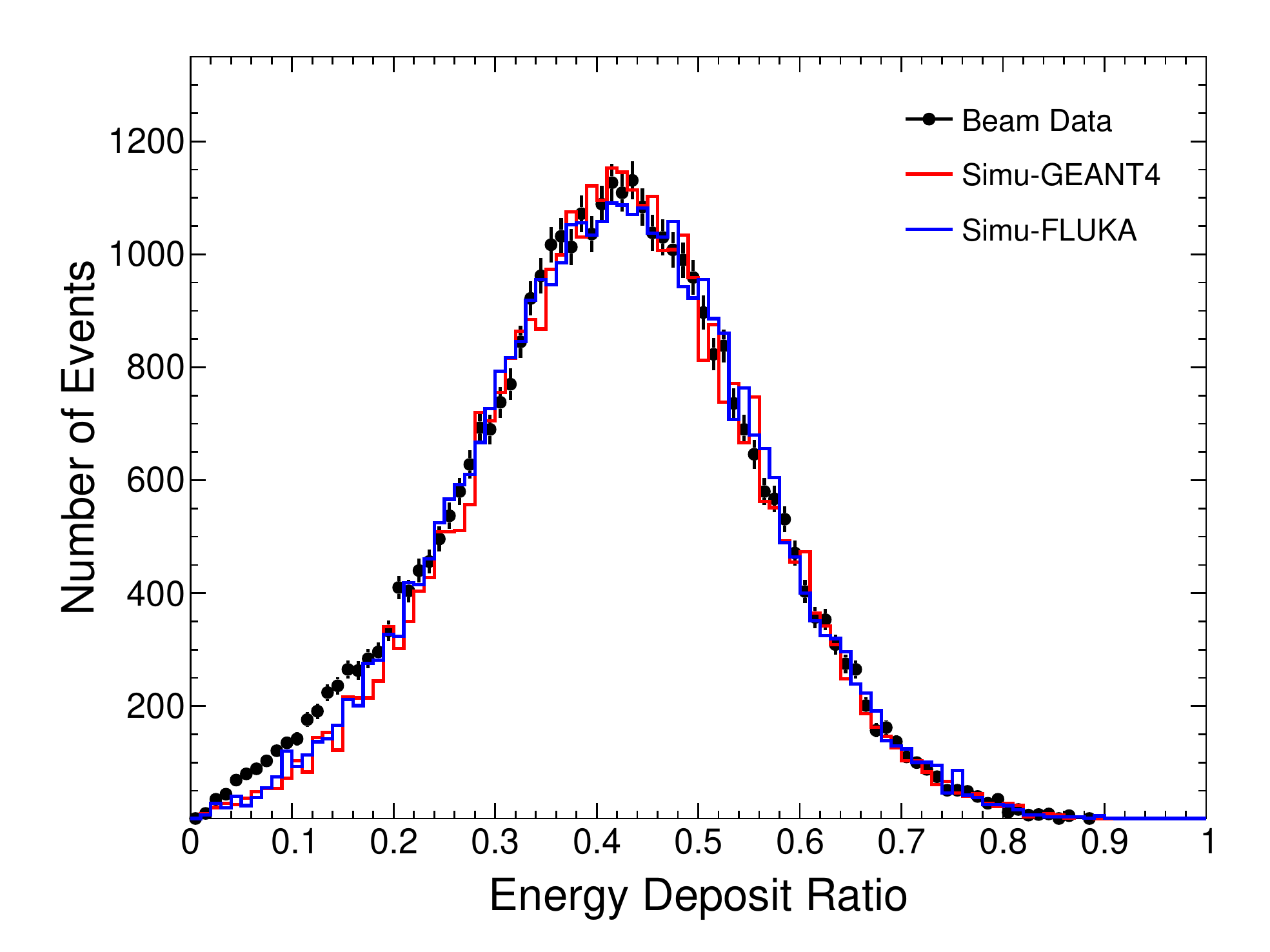}
  \includegraphics[width=.48\textwidth]{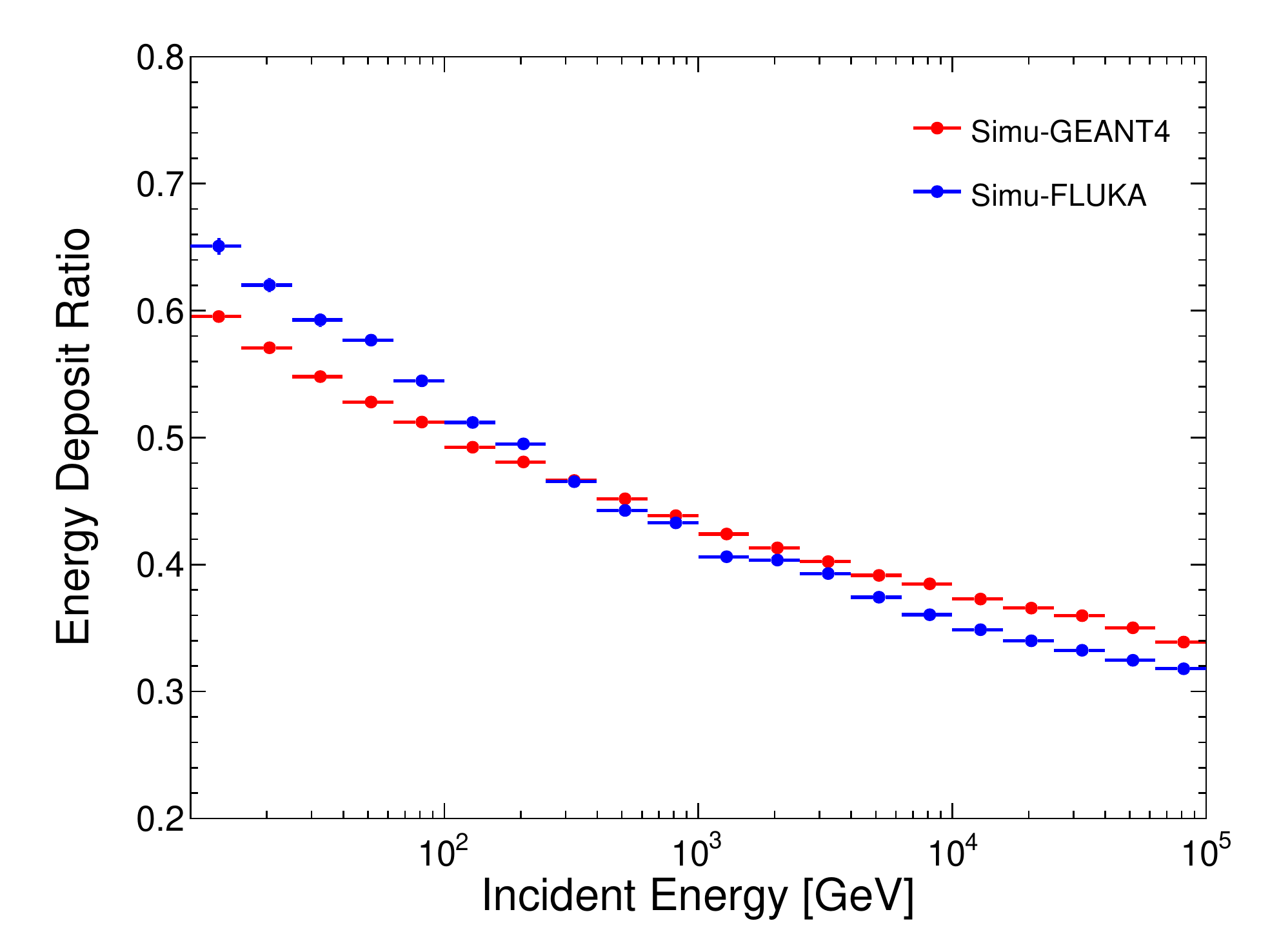}
  \caption{{\bf The energy response for protons from \GEANT and FLUKA.} 
  The top panel shows the distribution of the ratio of total energy deposit with respect to the incident energy
  for on-axis incident proton beams with 400 GeV/c momentum.
  Black, red and blue histograms correspond to Beam Data, \GEANT and FLUKA, respectively. 
  The bottom panel shows the most probable values of the energy deposit ratios as functions of incident energies for \GEANT (red) and FLUKA (blue), for an isotropic proton source with an $E^{-2.7}$ power-law spectrum.}
  \label{fig::edep}
\end{figure}

Different hadronic interaction models would present different energy response matrices \cite{DmpProton}, thereby leading to different deconvolutions for the initial spectrum of cosmic-ray proton. 
Before launch, the Engineering Qualification Model of DAMPE was extensively tested using test beams at the European Organization for Nuclear Research (CERN) in 2014-2015.
To compare with the test beam data, we generate MC samples follwing closely the settings of the test beams, such as the incident energies, hit points, and directions. We also apply the same event selections to both the beam test data and MC data as those in the flight data analysis \cite{DmpProton}, including the HE trigger, the track selection, the geometric cut, and the charge selection. The energy response of DAMPE for the on-axis incident proton beam with the momenta of 400 GeV/c is compared with the results from \GEANT and FLUKA simulations (see the top panel of Fig. \ref{fig::edep}).
Both \GEANT and FLUKA achieve good agreements with the beam test data, specifically at the momenta of 400 GeV/c.
To further compare the energy responses from \GEANT and FLUKA in the entire energy range of interest, an isotropic proton source with $E^{-1.0}$ spectrum from 10 GeV to 100 TeV is generated for the simulations. In the analysis, the spectra are re-weighted to $E^{-2.7}$ to be consistence with the CR flux.
The most probable values of the deposited energies obtained by fitting the energy ratio probabilities with an asymmetric gaussian function, along with the incident energy, are shown in the bottom panel of Fig. \ref{fig::edep}.
The energy responses of \GEANT  and FLUKA show an energy dependent difference from 10 GeV to 100 TeV, in consequence, the deconvoluted proton spectra based on \GEANT and FLUKA should have different spectral indices.

\subsection{Longitudinal development}

\begin{figure*}[htbp]
  \centering
  \includegraphics[width=.4\textwidth]{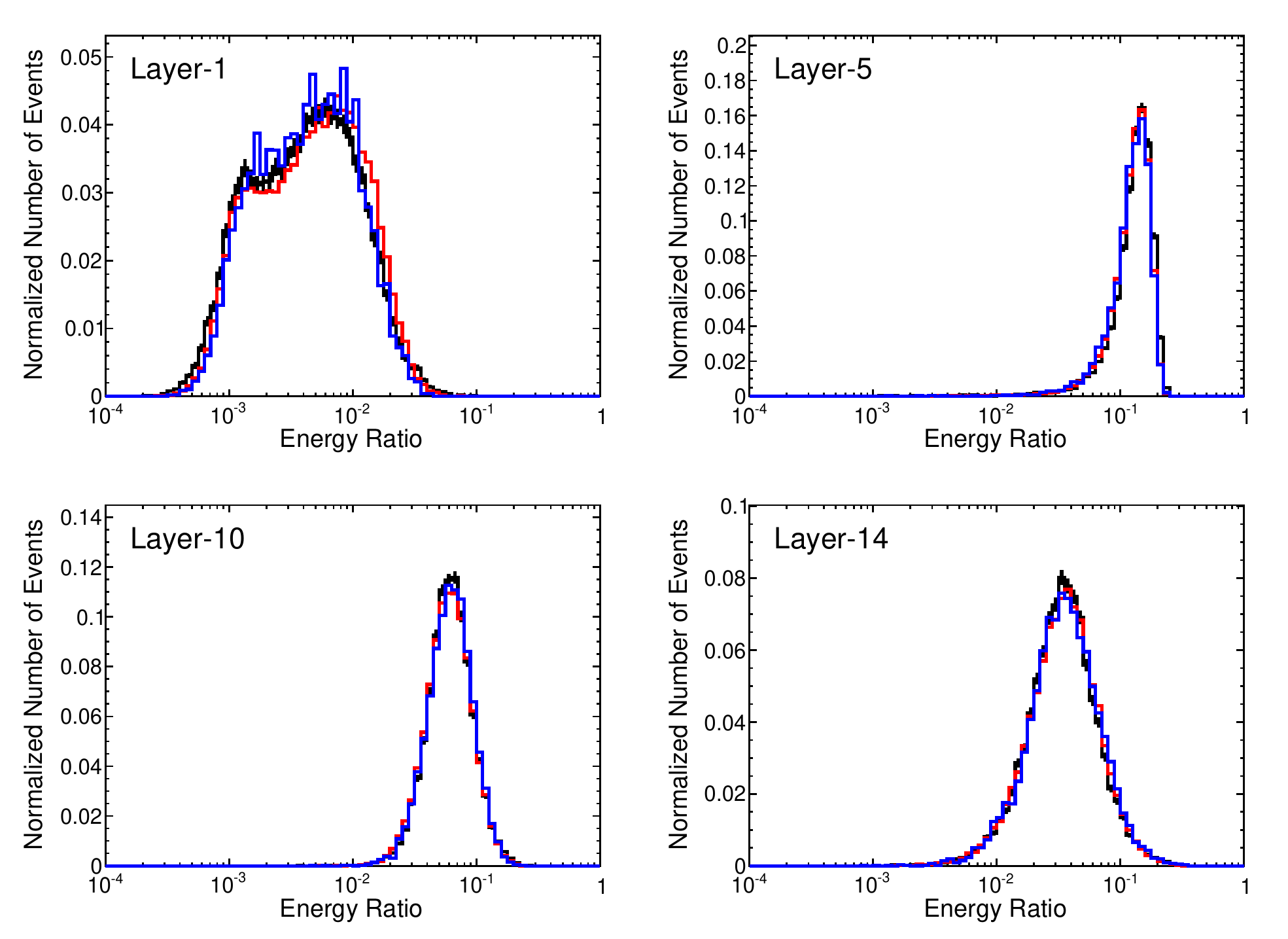}
  \includegraphics[width=.4\textwidth]{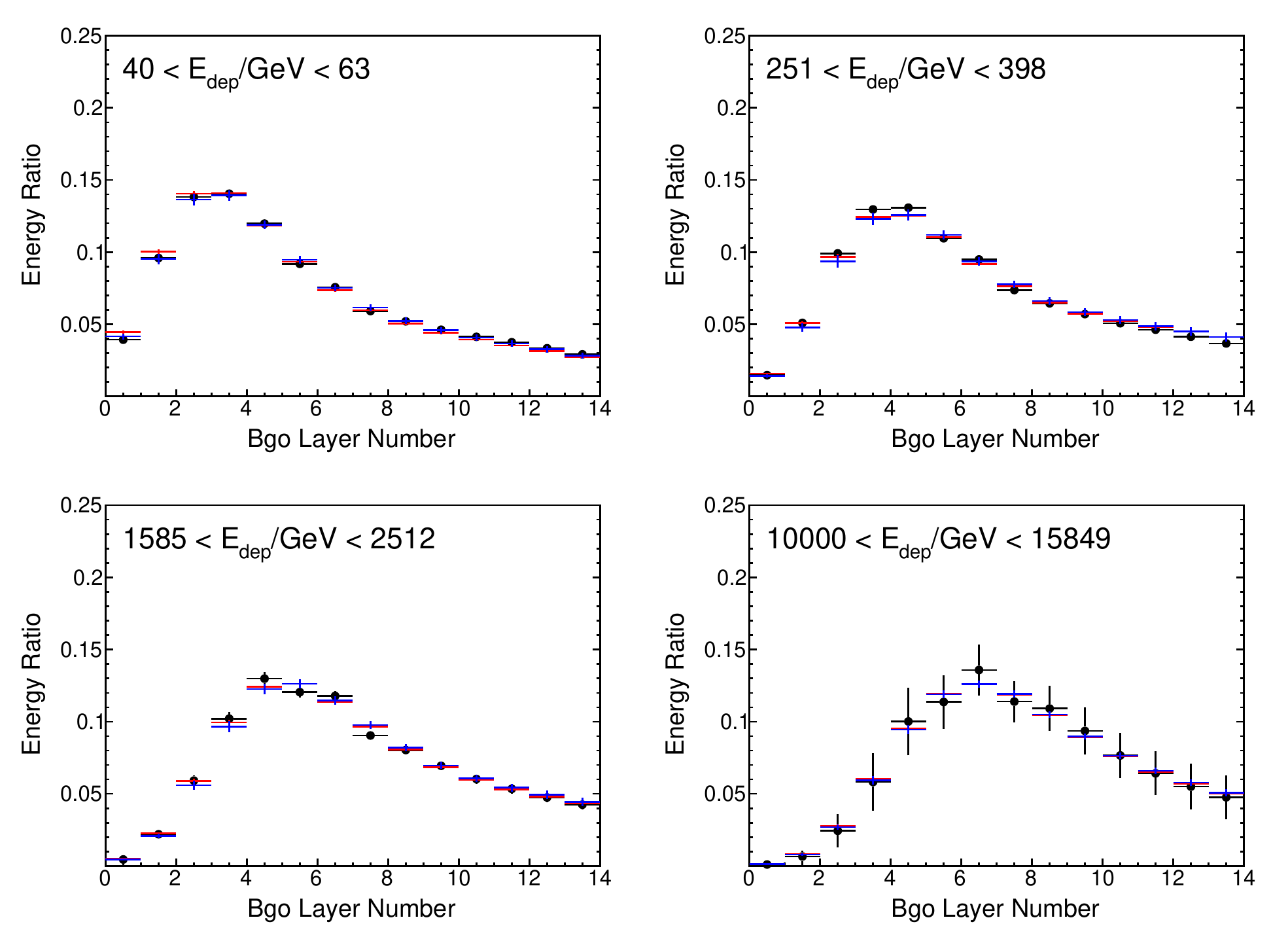}
  \caption{{\bf The longitudinal shower development for protons from \GEANT and FLUKA.} 
  The 4 plots on the left side show the energy ratio distributions in 4 typical BGO layers for total energy deposit between 1000 GeV and 1580 GeV.
  Black, red and blue histograms correspond to flight data, \GEANT and FLUKA, respectively.
  The 4 plots on the right side show the profiles of layer energy ratio in 4 typical energy deposit ranges. }
  \label{fig::eratio}
\end{figure*}

The longitudinal development of a hadronic shower is highly determined by the first inelastic interaction point, i.e. the inelastic scattering cross-section between the incident proton and the detector material. We calculate the ratios of the energy deposits in different BGO layers with the total energy deposit to describe the longitudinal shower development. 
Fig.~\ref{fig::eratio} shows the comparisons of layer energy ratios among flight data, \GEANT and FLUKA. 
While the DAMPE calorimeter is the thickest one in space, 
still the hadronic showers at these energies cannot be fully 
contained and a proportion of energy leaks in the bottom, 
as shown in the 4 plots on the right side of Fig.~\ref{fig::eratio}.
Both simulations (\GEANT and FLUKA) show good agreements of 
the longitudinal shower development with the flight data.

\subsection{Transverse development}

\begin{figure*}[htbp]
  \centering
  \includegraphics[width=.4\textwidth]{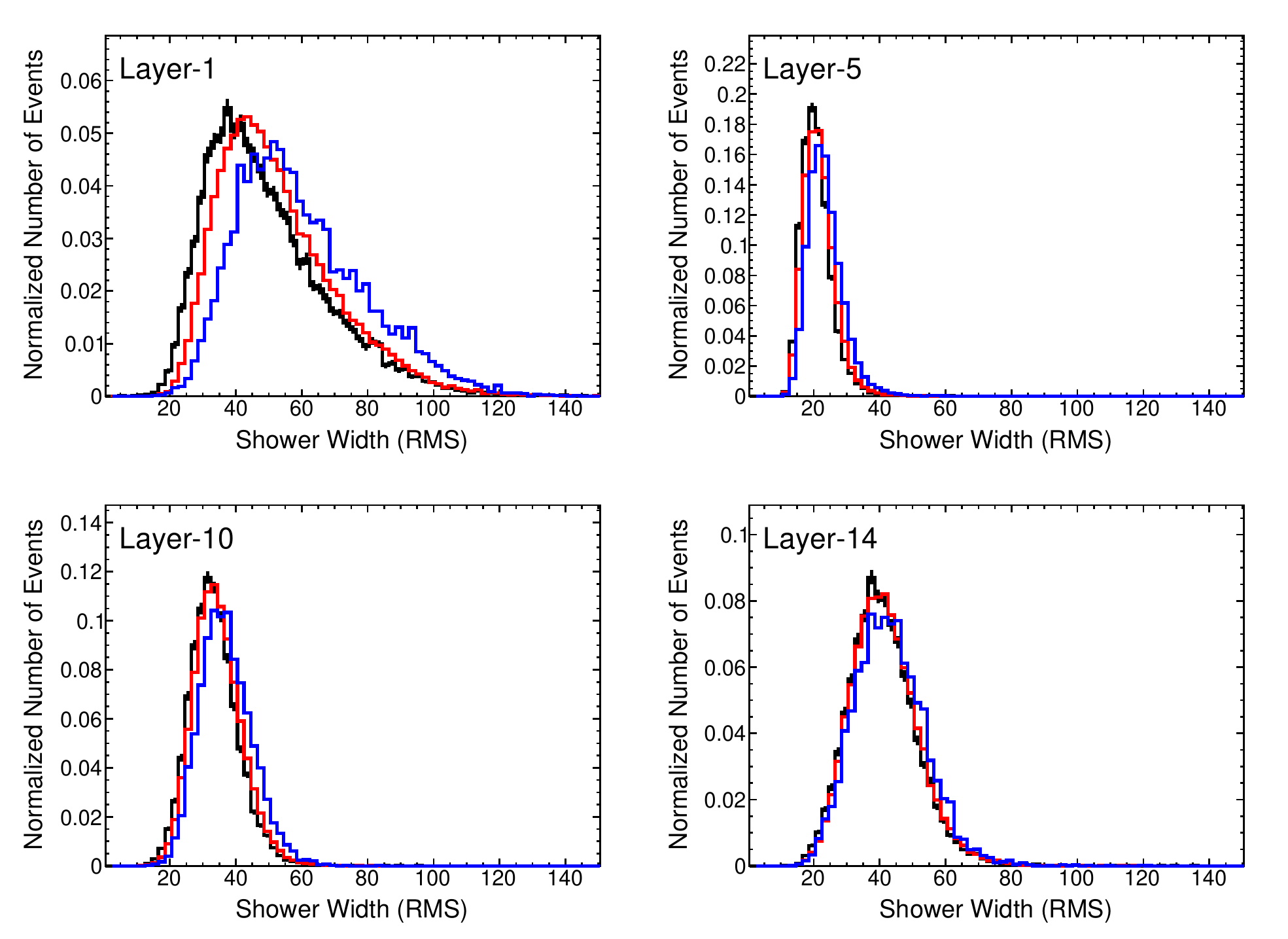}
  \includegraphics[width=.4\textwidth]{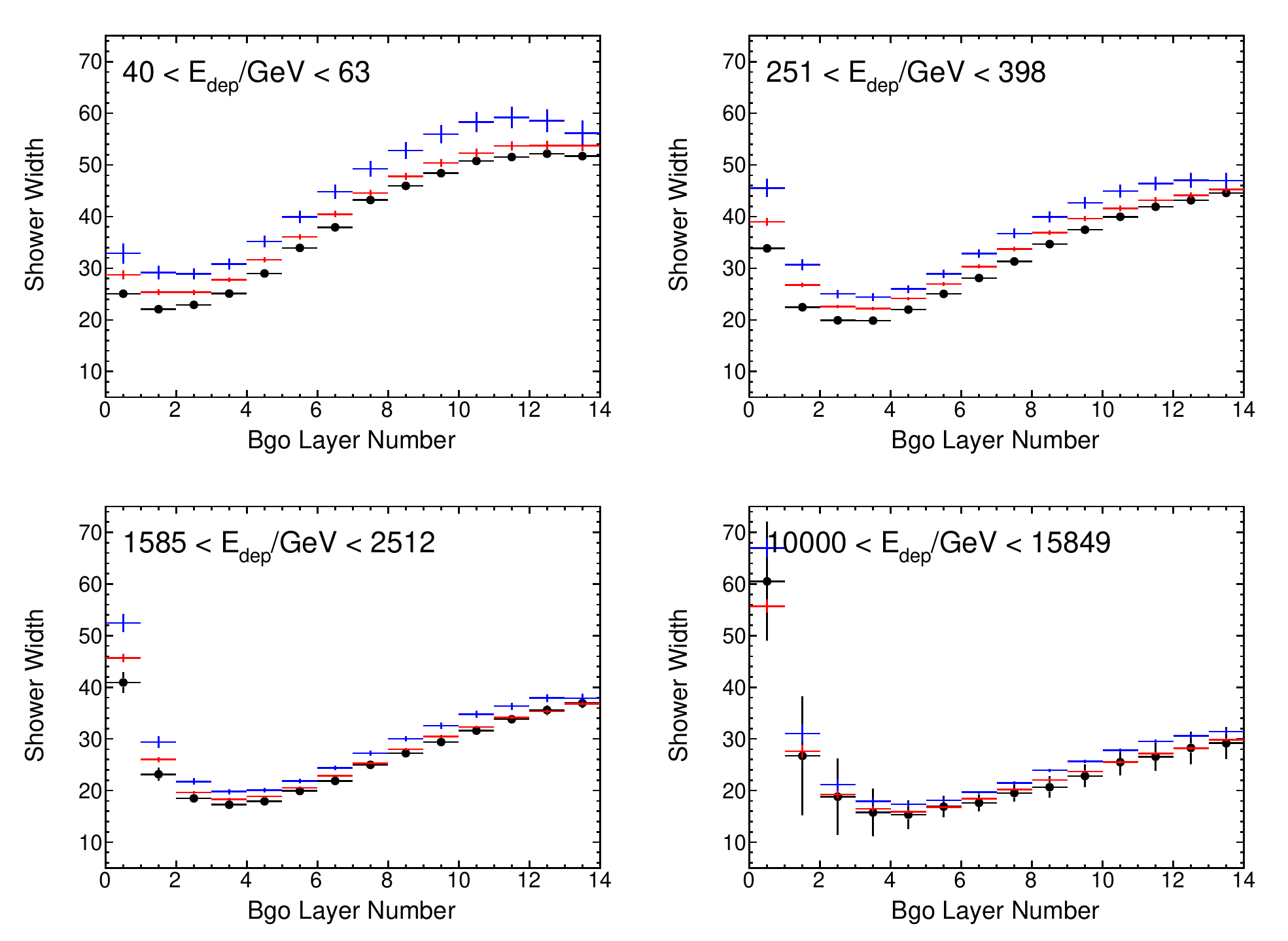}
  \caption{{\bf The transverse shower development for protons from \GEANT and FLUKA.} 
  The 4 plots on the left side show the RMS distributions in 4 typical BGO layers for total energy deposit between 1000 GeV and 1580 GeV.
  Black, red and blue histograms correspond to Flight data, \GEANT and FLUKA, respectively.
  The 4 plots on the right side show the profiles of layer energy ratio in 4 typical energy deposit ranges.}
  \label{fig::rms}
\end{figure*}

The transverse shower development, however, is intimately associated with the distribution of the types of subsidiary particles created through the interactions. We calculate the shower spread to characterize the transverse development, expressed by the energy-weighted root-mean-square (RMS) value of hit positions in the calorimeter. The RMS value of the fired $i$th layer is calculated as:
\begin{equation}
RMS_{i} = \sqrt{\frac{\Sigma_{j}(x_{j,i}-x_{c,i})^{2}E_{j,i}}{\Sigma_{j}E_{j,i}}}
\label{eq-RMS}
\end{equation}
Where $x_{j,i}$ and $E_{j,i}$ are the coordinates and energy deposit of the $j$th bar in the $i$th layer, and $x_{c,i}$ is the energy-weighted centre coordinate of the $i$th layer.  
Fig.~\ref{fig::rms} show the comparisons of RMS values in different layers among flight data, \GEANT and FLUKA. 
The differences among the FLUKA, \GEANT and flight data suggest some systematical uncertainties.
Overall, the results of \GEANT show  a better agreement with the flight data than the FULKA.
Based on the comparisons, we conclude that \GEANT carries out a more reliable simulation for the transverse development of the proton shower.

\subsection{Effect on the proton spectrum}
The absolute proton flux $F$ in an incident energy bin
$[E_i,\,E_i+\Delta E_i]$ can be calculated as
\begin{equation}
F(E_i,E_i+\Delta E_i) = \frac{N_{{\rm inc},i}}{A_{{\rm eff},i}~\Delta E_i
~T_{\rm exp}};~~~
 N_{{\rm inc},i}=\sum_{j=1}^{n}\mathbf{M}_{ij}N_{{\rm dep},j}\,,
\label{eq-flux}
\end{equation}
where $N_{{\rm inc},i}$ is the number of events in the $i$th incident
energy bin, $N_{{\rm dep},j}$ is the number of events in the $j$th deposited
energy bin, $\mathbf{M}_{ij}$ is the response matrix, $A_{{\rm eff},i}$
is the effective acceptance, $\Delta E_i$ is the width of the energy bin,
and $T_{\rm exp}$ is the exposure time. $N_{{\rm inc},i}$ in each incident
energy bin can be obtained via the unfolding procedure based on the Bayes
theorem \cite{Bayes}. 

\begin{figure}[htbp]
  \centering
  \includegraphics[width=.5\textwidth]{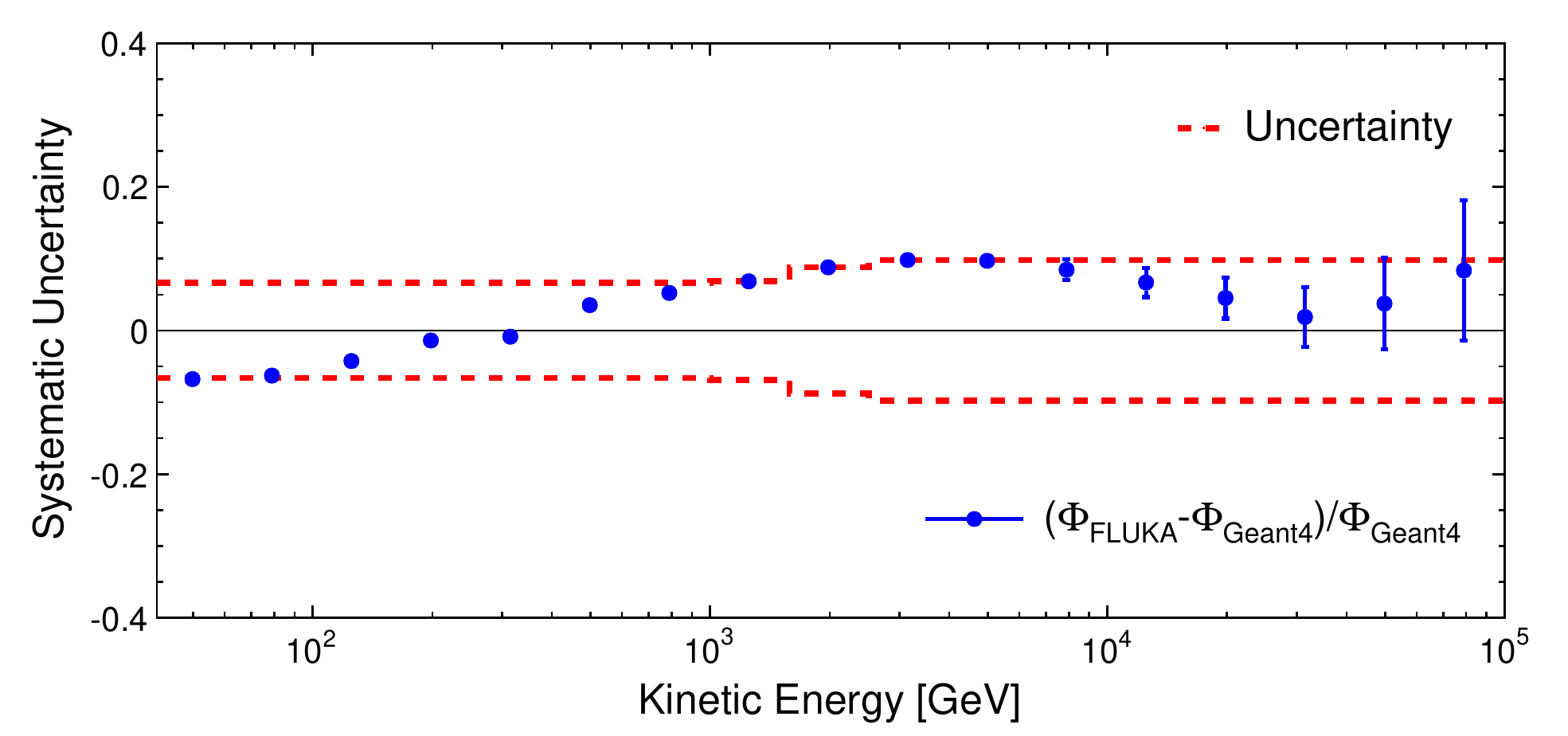}
  \caption{{\bf Energy dependence of the proton flux difference between \GEANT and FLUKA.}
  The blue points show the difference of measured proton spectrum assuming FLUKA simulation with respect to the spectrum based on \GEANT simulation.
  The dashed lines correspond to the associated systematic uncertainty claimed in Ref.~\cite{DmpProton}. }
  \label{fig::spectrum}
\end{figure}

The proton spectrum depends closely on the effective acceptance
and the energy response matrix, both are obtained from MC simulations.
The acceptance is obtained through calculating the fraction of events in
each incident energy bin survived from the whole selection procedure, and
the response matrix is obtained by counting the fraction of events in the
deposited energy bin $j$ for given incident energy bin $i$.
We applied the same selections as the flight data analysis \cite{DmpProton}
to obtain the corresponding effective acceptances and
energy response matrices for protons.
The effective acceptance from the FLUKA sample is lower than that from
the \GEANT sample by $\sim5\%$, which is dominated by the trigger
efficiency difference (see Fig.~\ref{fig::trigger}). On the other hand,
the energy response difference between two MC softwares
(see Fig.~\ref{fig::edep}) results in a complex effect on the fluxes
after the spectrum deconvolution.
The overall proton flux difference between \GEANT and FLUKA is shown in Fig.~\ref{fig::spectrum}.
Even though the maximum difference can be large as 10\%,
the global spectral structures are consistent with each other.
Based on the comparisons of shower development, we chose the \GEANT
spectrum as the benchmark, and take the difference between \GEANT
and FLUKA as the uncertainty. As shown in Fig.~\ref{fig::spectrum},
the proton flux difference varies from $-6.6\%$ to $9.8\%$, which is
taken as the systematic uncertainty due to different hadronic
interaction models \cite{DmpProton}.

 \section{Conclusion}
As a calorimeter-based experiment, DAMPE depends heavily on the precise
simulation of the interactions between the incident particle and the detector.
Due to the limited vertical thickness of the DAMPE calorimeter and the large
uncertainty for the hadronic interactions, the proton measurement is highly
associated with the simulation of the shower development. 
The comparison of the proton simulations of DAMPE between \GEANT and 
FLUKA has been carried out in this paper. We find that for given deposited
energies, these two simulations give basically similar results in describing 
the longitudinal developments of the proton showers in the BGO calorimeter.
The lateral distributions, however, show several differences. For the
overall energy deposition, the FLUKA results are higher by $(3\sim8)\%$
for primary energies below 1 TeV, and lower by $(2\sim5\%)$ above 1 TeV.
The shower developments also affect the trigger efficiency evaluation
of protons, which is leading to a deviation about $5\%$ between the results of these two simulation
softwares. The overall uncertainties due to the hadronic models are
estimated to be about $10\%$.

\vspace{3mm}
{\bf Acknowledgments}
This work is partly supported by the National Key Research and Development Program 
of China (Grant No. 2016YFA0400200), and the National Natural Science Foundation of China
(Grant Nos. 11722328, 11773085, U1738127, U1738138, U1738205, U1738207, 11851305),
the 100 Talents Program of Chinese Academy of Sciences,
the Youth Innovation Promotion Association CAS,
and the Program for Innovative Talents and Entrepreneur in Jiangsu.
In Europe the activities are supported by the Swiss National Science Foundation (SNSF), Switzerland,
and the National Institute for Nuclear Physics (INFN), Italy.

\bibliographystyle{apsrev}
\bibliography{dampe-simu}

\begin{thebibliography}{32}
\expandafter\ifx\csname natexlab\endcsname\relax\def\natexlab#1{#1}\fi
\expandafter\ifx\csname bibnamefont\endcsname\relax
  \def\bibnamefont#1{#1}\fi
\expandafter\ifx\csname bibfnamefont\endcsname\relax
  \def\bibfnamefont#1{#1}\fi
\expandafter\ifx\csname citenamefont\endcsname\relax
  \def\citenamefont#1{#1}\fi
\expandafter\ifx\csname url\endcsname\relax
  \def\url#1{\texttt{#1}}\fi
\expandafter\ifx\csname urlprefix\endcsname\relax\def\urlprefix{URL }\fi
\providecommand{\bibinfo}[2]{#2}
\providecommand{\eprint}[2][]{\url{#2}}

\bibitem[{\citenamefont{Adriani et~al.}(2011)}]{Adriani:2011cu}
\bibinfo{author}{\bibfnamefont{O.}~\bibnamefont{Adriani}} \bibnamefont{et~al.}
  (\bibinfo{collaboration}{PAMELA}), \bibinfo{journal}{Science}
  \textbf{\bibinfo{volume}{332}}, \bibinfo{pages}{69} (\bibinfo{year}{2011}),
  \eprint{1103.4055}.

\bibitem[{\citenamefont{Aguilar et~al.}(2015)}]{Aguilar:2015ooa}
\bibinfo{author}{\bibfnamefont{M.}~\bibnamefont{Aguilar}} \bibnamefont{et~al.}
  (\bibinfo{collaboration}{AMS}), \bibinfo{journal}{Phys. Rev. Lett.}
  \textbf{\bibinfo{volume}{114}}, \bibinfo{pages}{171103}
  (\bibinfo{year}{2015}).

\bibitem[{\citenamefont{Aguilar et~al.}(2017)}]{Aguilar:2017hno}
\bibinfo{author}{\bibfnamefont{M.}~\bibnamefont{Aguilar}} \bibnamefont{et~al.}
  (\bibinfo{collaboration}{AMS}), \bibinfo{journal}{Phys. Rev. Lett.}
  \textbf{\bibinfo{volume}{119}}, \bibinfo{pages}{251101}
  (\bibinfo{year}{2017}).

\bibitem[{\citenamefont{Tanabashi et~al.}(2018)}]{Tanabashi:2018oca}
\bibinfo{author}{\bibfnamefont{M.}~\bibnamefont{Tanabashi}}
  \bibnamefont{et~al.} (\bibinfo{collaboration}{Particle Data Group}),
  \bibinfo{journal}{Phys. Rev. D} \textbf{\bibinfo{volume}{98}},
  \bibinfo{pages}{030001} (\bibinfo{year}{2018}).

\bibitem[{\citenamefont{Torii and Marrocchesi}(2019)}]{Torii:2019wxn}
\bibinfo{author}{\bibfnamefont{S.}~\bibnamefont{Torii}} \bibnamefont{and}
  \bibinfo{author}{\bibfnamefont{P.~S.} \bibnamefont{Marrocchesi}}
  (\bibinfo{collaboration}{CALET}), \bibinfo{journal}{Adv. Space Res.}
  \textbf{\bibinfo{volume}{64}}, \bibinfo{pages}{2531} (\bibinfo{year}{2019}).

\bibitem[{\citenamefont{Atkin et~al.}(2015)}]{Atkin:2015nrd}
\bibinfo{author}{\bibfnamefont{E.}~\bibnamefont{Atkin}} \bibnamefont{et~al.},
  \bibinfo{journal}{EPJ Web Conf.} \textbf{\bibinfo{volume}{105}},
  \bibinfo{pages}{01002} (\bibinfo{year}{2015}).

\bibitem[{\citenamefont{{Chang}}(2014)}]{ChangJ2014}
\bibinfo{author}{\bibfnamefont{J.}~\bibnamefont{{Chang}}},
  \bibinfo{journal}{Chinese Journal of Space Science}
  \textbf{\bibinfo{volume}{34}}, \bibinfo{pages}{550} (\bibinfo{year}{2014}).

\bibitem[{\citenamefont{{Chang} et~al.}(2017)}]{DmpMission}
\bibinfo{author}{\bibfnamefont{J.}~\bibnamefont{{Chang}}} \bibnamefont{et~al.}
  (\bibinfo{collaboration}{DAMPE}), \bibinfo{journal}{Astroparticle Physics}
  \textbf{\bibinfo{volume}{95}}, \bibinfo{pages}{6} (\bibinfo{year}{2017}).

\bibitem[{\citenamefont{Kang et~al.}(2019)}]{Kang:2019pfl}
\bibinfo{author}{\bibfnamefont{S.}~\bibnamefont{Kang}} \bibnamefont{et~al.},
  \bibinfo{journal}{Adv. Space Res.} \textbf{\bibinfo{volume}{64}},
  \bibinfo{pages}{2564} (\bibinfo{year}{2019}).

\bibitem[{\citenamefont{Yuan and Feng}(2018)}]{Yuan:2018rys}
\bibinfo{author}{\bibfnamefont{Q.}~\bibnamefont{Yuan}} \bibnamefont{and}
  \bibinfo{author}{\bibfnamefont{L.}~\bibnamefont{Feng}},
  \bibinfo{journal}{Sci. China Phys. Mech. Astron.}
  \textbf{\bibinfo{volume}{61}}, \bibinfo{pages}{101002}
  (\bibinfo{year}{2018}), \eprint{1807.11638}.

\bibitem[{\citenamefont{Yu et~al.}(2017)}]{YuYH2017}
\bibinfo{author}{\bibfnamefont{Y.}~\bibnamefont{Yu}} \bibnamefont{et~al.},
  \bibinfo{journal}{Astroparticle Physics} \textbf{\bibinfo{volume}{94}},
  \bibinfo{pages}{1} (\bibinfo{year}{2017}), \eprint{1703.00098}.

\bibitem[{\citenamefont{Azzarello et~al.}(2016)}]{Azzarello2016}
\bibinfo{author}{\bibfnamefont{P.}~\bibnamefont{Azzarello}}
  \bibnamefont{et~al.}, \bibinfo{journal}{Nuclear Instruments and Methods in
  Physics Research A} \textbf{\bibinfo{volume}{831}}, \bibinfo{pages}{378}
  (\bibinfo{year}{2016}).

\bibitem[{\citenamefont{{Zhang} et~al.}(2016)}]{ZhangZY2016}
\bibinfo{author}{\bibfnamefont{Z.}~\bibnamefont{{Zhang}}} \bibnamefont{et~al.},
  \bibinfo{journal}{Nuclear Instruments and Methods in Physics Research A}
  \textbf{\bibinfo{volume}{836}}, \bibinfo{pages}{98} (\bibinfo{year}{2016}).

\bibitem[{\citenamefont{{Huang} et~al.}(2020)\citenamefont{{Huang}, {Ma},
  {Yue}, {Zhang}, {Chang}, {Dong}, and {Zhang}}}]{HuangYY2020}
\bibinfo{author}{\bibfnamefont{Y.-Y.} \bibnamefont{{Huang}}},
  \bibinfo{author}{\bibfnamefont{T.}~\bibnamefont{{Ma}}},
  \bibinfo{author}{\bibfnamefont{C.}~\bibnamefont{{Yue}}},
  \bibinfo{author}{\bibfnamefont{Y.}~\bibnamefont{{Zhang}}},
  \bibinfo{author}{\bibfnamefont{J.}~\bibnamefont{{Chang}}},
  \bibinfo{author}{\bibfnamefont{T.-K.} \bibnamefont{{Dong}}},
  \bibnamefont{and} \bibinfo{author}{\bibfnamefont{Y.-Q.}
  \bibnamefont{{Zhang}}}, \bibinfo{journal}{Research in Astronomy and
  Astrophysics (in press)}  (\bibinfo{year}{2020}), \eprint{2005.07828}.

\bibitem[{\citenamefont{Ambrosi et~al.}(2019)}]{DmpCalibration}
\bibinfo{author}{\bibfnamefont{G.}~\bibnamefont{Ambrosi}} \bibnamefont{et~al.}
  (\bibinfo{collaboration}{DAMPE}), \bibinfo{journal}{Astroparticle Physics}
  \textbf{\bibinfo{volume}{106}}, \bibinfo{pages}{18} (\bibinfo{year}{2019}),
  \eprint{1907.02173}.

\bibitem[{\citenamefont{Ambrosi et~al.}(2017)}]{DmpElectron}
\bibinfo{author}{\bibfnamefont{G.}~\bibnamefont{Ambrosi}} \bibnamefont{et~al.}
  (\bibinfo{collaboration}{DAMPE}), \bibinfo{journal}{Nature}
  \textbf{\bibinfo{volume}{552}}, \bibinfo{pages}{63} (\bibinfo{year}{2017}).

\bibitem[{\citenamefont{An et~al.}(2019)}]{DmpProton}
\bibinfo{author}{\bibfnamefont{Q.}~\bibnamefont{An}} \bibnamefont{et~al.}
  (\bibinfo{collaboration}{DAMPE}), \bibinfo{journal}{Science Advances}
  \textbf{\bibinfo{volume}{5}}, \bibinfo{pages}{eaax3793}
  (\bibinfo{year}{2019}), \eprint{1909.12860}.

\bibitem[{\citenamefont{Agostinelli et~al.}(2003)\citenamefont{Agostinelli,
  Allison, Amako et~al.}}]{GEANT4}
\bibinfo{author}{\bibfnamefont{S.}~\bibnamefont{Agostinelli}},
  \bibinfo{author}{\bibfnamefont{J.}~\bibnamefont{Allison}},
  \bibinfo{author}{\bibfnamefont{K.}~\bibnamefont{Amako}},
  \bibnamefont{et~al.}, \bibinfo{journal}{Nuclear Instruments and Methods in
  Physics Research A} \textbf{\bibinfo{volume}{506}}, \bibinfo{pages}{250}
  (\bibinfo{year}{2003}).

\bibitem[{\citenamefont{Ferrari et~al.}(2005)\citenamefont{Ferrari, Sala,
  Fass{\"o}, and Ranft}}]{FLUKA2005}
\bibinfo{author}{\bibfnamefont{A.}~\bibnamefont{Ferrari}},
  \bibinfo{author}{\bibfnamefont{P.}~\bibnamefont{Sala}},
  \bibinfo{author}{\bibfnamefont{A.}~\bibnamefont{Fass{\"o}}},
  \bibnamefont{and} \bibinfo{author}{\bibfnamefont{J.}~\bibnamefont{Ranft}},
  \emph{\bibinfo{title}{{FLUKA : A multi-particle transport code}}}, CERN
  Yellow Reports: Monographs (\bibinfo{publisher}{CERN},
  \bibinfo{address}{Geneva}, \bibinfo{year}{2005}), ISBN
  \bibinfo{isbn}{9290832606},
  \urlprefix\url{https://cds.cern.ch/record/898301}.

\bibitem[{\citenamefont{B{\"o}hlen et~al.}(2014)\citenamefont{B{\"o}hlen,
  Cerutti, Chin et~al.}}]{FLUKA}
\bibinfo{author}{\bibfnamefont{T.~T.} \bibnamefont{B{\"o}hlen}},
  \bibinfo{author}{\bibfnamefont{F.}~\bibnamefont{Cerutti}},
  \bibinfo{author}{\bibfnamefont{M.~P.~W.} \bibnamefont{Chin}},
  \bibnamefont{et~al.} (\bibinfo{collaboration}{FLUKA Collaboration}),
  \bibinfo{journal}{Nuclear Data Sheets} \textbf{\bibinfo{volume}{120}},
  \bibinfo{pages}{211} (\bibinfo{year}{2014}).

\bibitem[{\citenamefont{{Bolshakova} et~al.}(2010)\citenamefont{{Bolshakova},
  {Boyko}, {Chelkov} et~al.}}]{ComparisonGF}
\bibinfo{author}{\bibfnamefont{A.}~\bibnamefont{{Bolshakova}}},
  \bibinfo{author}{\bibfnamefont{I.}~\bibnamefont{{Boyko}}},
  \bibinfo{author}{\bibfnamefont{G.}~\bibnamefont{{Chelkov}}},
  \bibnamefont{et~al.}, \bibinfo{journal}{European Physical Journal C}
  \textbf{\bibinfo{volume}{70}}, \bibinfo{pages}{543} (\bibinfo{year}{2010}),
  \eprint{1006.3429}.

\bibitem[{\citenamefont{{Allison} et~al.}(2016)\citenamefont{{Allison},
  {Amako}, {Apostolakis}, {Arce} et~al.}}]{GEANT4_2016}
\bibinfo{author}{\bibfnamefont{J.}~\bibnamefont{{Allison}}},
  \bibinfo{author}{\bibfnamefont{K.}~\bibnamefont{{Amako}}},
  \bibinfo{author}{\bibfnamefont{J.}~\bibnamefont{{Apostolakis}}},
  \bibinfo{author}{\bibfnamefont{P.}~\bibnamefont{{Arce}}},
  \bibnamefont{et~al.}, \bibinfo{journal}{Nuclear Instruments and Methods in
  Physics Research A} \textbf{\bibinfo{volume}{835}}, \bibinfo{pages}{186}
  (\bibinfo{year}{2016}).

\bibitem[{\citenamefont{Andersson et~al.}(1987)\citenamefont{Andersson,
  Gustafson, and Nilsson-Almqvist}}]{FTFP_G4_1}
\bibinfo{author}{\bibfnamefont{B.}~\bibnamefont{Andersson}},
  \bibinfo{author}{\bibfnamefont{G.}~\bibnamefont{Gustafson}},
  \bibnamefont{and}
  \bibinfo{author}{\bibfnamefont{B.}~\bibnamefont{Nilsson-Almqvist}},
  \bibinfo{journal}{Nuclear Physics B} \textbf{\bibinfo{volume}{281}},
  \bibinfo{pages}{289 } (\bibinfo{year}{1987}), ISSN \bibinfo{issn}{0550-3213}.

\bibitem[{\citenamefont{Nilsson-Almqvist and Stenlund}(1987)}]{FTFP_G4_2}
\bibinfo{author}{\bibfnamefont{B.}~\bibnamefont{Nilsson-Almqvist}}
  \bibnamefont{and} \bibinfo{author}{\bibfnamefont{E.}~\bibnamefont{Stenlund}},
  \bibinfo{journal}{Computer Physics Communications}
  \textbf{\bibinfo{volume}{43}}, \bibinfo{pages}{387 } (\bibinfo{year}{1987}),
  ISSN \bibinfo{issn}{0010-4655}.

\bibitem[{\citenamefont{GEANT4 Collaboration}({\natexlab{a}})}]{GEANT4_PhLG}
\bibinfo{author}{\bibnamefont{GEANT4 Collaboration}},
  \emph{\bibinfo{title}{Geant4 physics list guide}},
  \bibinfo{howpublished}{\url{http://geant4-userdoc.web.cern.ch/geant4-userdoc/UsersGuides/PhysicsListGuide/html/index.html}}.

\bibitem[{\citenamefont{Roesler et~al.}(2001)\citenamefont{Roesler, Engel, and
  Ranft}}]{dpmjetfluka}
\bibinfo{author}{\bibfnamefont{S.}~\bibnamefont{Roesler}},
  \bibinfo{author}{\bibfnamefont{R.}~\bibnamefont{Engel}}, \bibnamefont{and}
  \bibinfo{author}{\bibfnamefont{J.}~\bibnamefont{Ranft}},
  \bibinfo{journal}{Advanced Monte Carlo for Radiation Physics, Particle
  Transport Simulation and Applications} pp. \bibinfo{pages}{1033–--1038}
  (\bibinfo{year}{2001}).

\bibitem[{\citenamefont{{Chytracek} et~al.}(2006)\citenamefont{{Chytracek},
  {Mccormick}, {Pokorski}, and {Santin}}}]{GDML}
\bibinfo{author}{\bibfnamefont{R.}~\bibnamefont{{Chytracek}}},
  \bibinfo{author}{\bibfnamefont{J.}~\bibnamefont{{Mccormick}}},
  \bibinfo{author}{\bibfnamefont{W.}~\bibnamefont{{Pokorski}}},
  \bibnamefont{and} \bibinfo{author}{\bibfnamefont{G.}~\bibnamefont{{Santin}}},
  \bibinfo{journal}{IEEE Transactions on Nuclear Science}
  \textbf{\bibinfo{volume}{53}}, \bibinfo{pages}{2892} (\bibinfo{year}{2006}).

\bibitem[{\citenamefont{Wang et~al.}(2017)}]{DMPSW}
\bibinfo{author}{\bibfnamefont{C.}~\bibnamefont{Wang}} \bibnamefont{et~al.},
  \bibinfo{journal}{Chinese Physics C} \textbf{\bibinfo{volume}{41}},
  \bibinfo{pages}{106201} (\bibinfo{year}{2017}).

\bibitem[{\citenamefont{Emmett}(1975)}]{MORSE}
\bibinfo{author}{\bibfnamefont{M.~B.} \bibnamefont{Emmett}},
  \bibinfo{type}{Tech. Rep.} \bibinfo{number}{ORNL-4972},
  \bibinfo{institution}{Oak Ridge National Laboratory, United States}
  (\bibinfo{year}{1975}).

\bibitem[{\citenamefont{GEANT4 Collaboration}({\natexlab{b}})}]{GEANT4_PhRM}
\bibinfo{author}{\bibnamefont{GEANT4 Collaboration}},
  \emph{\bibinfo{title}{Geant4 physics reference manual}},
  \bibinfo{howpublished}{\url{https://geant4-userdoc.web.cern.ch/geant4-userdoc/UsersGuides/PhysicsReferenceManual/html/index.html}}.

\bibitem[{\citenamefont{Zhang et~al.}(2019)\citenamefont{Zhang, Guo, Liu
  et~al.}}]{ZhangYQ2019}
\bibinfo{author}{\bibfnamefont{Y.-Q.} \bibnamefont{Zhang}},
  \bibinfo{author}{\bibfnamefont{J.-H.} \bibnamefont{Guo}},
  \bibinfo{author}{\bibfnamefont{Y.}~\bibnamefont{Liu}}, \bibnamefont{et~al.},
  \bibinfo{journal}{Research in Astronomy and Astrophysics (RAA)}
  \textbf{\bibinfo{volume}{19}}, \bibinfo{pages}{123} (\bibinfo{year}{2019}),
  ISSN \bibinfo{issn}{2397-6209}.

\bibitem[{\citenamefont{{D'Agostini}}(1995)}]{Bayes}
\bibinfo{author}{\bibfnamefont{G.}~\bibnamefont{{D'Agostini}}},
  \bibinfo{journal}{Nuclear Instruments and Methods in Physics Research A}
  \textbf{\bibinfo{volume}{362}}, \bibinfo{pages}{487} (\bibinfo{year}{1995}).

\end{thebibliography}
\end{document}